\documentstyle[12pt,fleqn,epsf]{article}
\begin{document}
\begin{flushright}
APCTP-97-12\\
SNUTP-97-065\\
hep-th/9707038
\end{flushright}
\begin{center}
{\Large\bf Finite Energy Electroweak Monopoles}
\vskip 2em
{\large\bf Y. M. Cho}\\
{Asia Pacific Center for Theoretical Physics
\\and
\\ Department of Physics, Seoul National University, Seoul 151-742, Korea}\\
\vskip 2em
{\large\bf Kyoungtae Kimm}\\
{Center for Theoretical Physics, Seoul National University, Seoul 151-742,
Korea}
\vskip 2em
{\bf Abstract}
\end{center}
\begin{quote}
We present finite 
energy analytic monopole and dyon
solutions whose size is fixed by the electroweak scale.
The new solutions are obtained by regularizing the recent
Cho-Maison solutions of the Weinberg-Salam model.
Our result shows that genuine 
electroweak monopole and dyon could exist whose
mass scale is much smaller than the grand unification scale.
\end{quote}

\setcounter{section}{1}
\setcounter{equation}{0}
\section*{I. Introduction}
\indent 

Ever since Dirac \cite{Dirac} has introduced the concept of the
magnetic monopole, the monopoles have remained a fascinating subject 
in theoretical physics.
The Abelian monopole has been generalized to the non-Abelian 
gauge theory by Wu and Yang \cite{Wu} who showed that
the pure $SU(2)$ gauge theory allows a point-like monopole,
and by 't Hooft and Polyakov \cite{Hooft} who have constructed
a finite energy monopole solution in the Georgi-Glashow model
as a topological soliton. 
In the interesting case of the electroweak theory of Weinberg and
Salam, however, it has generally been believed that there
exists no topological monopole of physical interest.
The basis for this ``non-existence theorem'' is, of course, that with
the spontaneous symmetry breaking the quotient space $SU(2) \times
U(1)/U(1)_{\rm em}$ allows no non-trivial second homotopy.
This has led many people  to conclude that 
there is no topological structure in
the Weinberg-Salam model which can accommodate a magnetic monopole.

This, however, is shown to be not true. Indeed recently
Cho and Maison~\cite{Cho0} have established 
that the Weinberg-Salam model has exactly the same topological
structure as the Georgi-Glashow model which allows the
magnetic monopoles, and demonstrated the existence of a new type of 
monopole and dyon solutions in the standard Weinberg-Salam
model. 
{\em This was based on the observation that 
the Weinberg-Salam model, with the hypercharge $U(1)$,
could be viewed as a gauged $CP^1$ model in which the (normalized)
Higgs doublet plays the role of the $CP^1$ field.} So 
the Weinberg-Salam model does have exactly the same nontrivial
second homotopy as the Georgi-Glashow model which allows topological monopoles.
Once this is understood one could proceed to construct the desired monopole
and dyon solutions in the Weinberg-Salam model.
Originally the Cho-Maison solutions were obtained by a numerical integration, 
but now a mathematically
rigorous existence proof has been established which supports
the numerical results \cite{Yang}.

The Cho-Maison monopole may be viewed as a hybrid between the Dirac monopole
and the 't Hooft-Polyakov monopole, because it has a $U(1)$ point
singularity at the center even though the $SU(2)$ part is completely
regular. Consequently it carries an infinite energy so that 
at the classical level the mass of the monopole remains arbitrary. 
{\em A priori} there is nothing
wrong with this, but nevertheless one
may wonder whether one can have an analytic 
electroweak monopole which has a finite 
energy.
The purpose of this paper is to show that this is indeed possible,
and to present explicit electroweak monopole solutions with finite energy.
Clearly the new monopoles should have important physical
applications in the phenomenology of electroweak interaction.

\setcounter{section}{2}
\setcounter{equation}{0}
\section*{II. Monopoles in Weinberg-Salam Model}
\indent

Before we construct the finite energy monopole solutions we must understand how 
one could obtain the infinite energy solutions first. So we will briefly
review the Cho-Maison solutions in the Weinberg-Salam
model.
Let us start with the Lagrangian which
describes (the bosonic sector of) the standard Weinberg-Salam model 
\begin{eqnarray}
{\cal L} &=& -|\hat{D}_{\mu}\mbox{\boldmath $\phi$}|^2 
 -\frac{\lambda}{2}\Big(\mbox{\boldmath $\phi$}^\dagger\mbox{\boldmath $\phi$}
 -\frac{\mu^2}{\lambda}\Big)^2
 -\frac{1}{4}(\mbox{\boldmath $ F$}_{\mu\nu})^2 
 -\frac{1}{4}(G_{\mu\nu})^2 , 
\label{lag1}
\end{eqnarray}
\begin{eqnarray}
\hat{D}_{\mu} \mbox{\boldmath $\mbox{\boldmath ${\phi}$}$}
&=& \Big(\partial_{\mu} 
  -i\frac{g}{2}\mbox{\boldmath $\tau$}\!\cdot\!\mbox{\boldmath $A$}_{\mu}
  -i\frac{g'}{2} B_{\mu}\Big) \mbox{\boldmath $\phi$}
  =\Big(D_{\mu}
        -i\frac{g'}{2}B_{\mu}\Big)\mbox{\boldmath $\phi$},\nonumber
\nonumber 
\end{eqnarray}
where ${\bf \mbox{\boldmath $\phi$}}$ is the Higgs doublet, 
$\mbox{\boldmath $F$}_{\mu\nu}$
and
$G_{\mu\nu}$ 
are the gauge field strengths of $SU(2)$ and $U(1)$ with the
potentials $\mbox{\boldmath $A$}_{\mu}$ and $B_{\mu}$, and $g$ and $g'$
are the corresponding coupling constants.
Notice that $D_{\mu}$ describes the covariant derivative  of the $SU(2)$
subgroup only. From (\ref{lag1}) one has the following equations of motion
\begin{eqnarray}
\hat{D}_{\mu}(\hat{D}_{\mu}\mbox{\boldmath $\phi$})
   &=&\lambda\Big(\mbox{\boldmath ${\phi}$}^{\dagger} 
                  \mbox{\boldmath ${\phi}$}- \frac {\mu^2}{\lambda} \Big) 
                  \mbox{\boldmath ${\phi}$},
\nonumber
\end{eqnarray}
\begin{eqnarray}
D_{\mu} \mbox{\boldmath $F$}_{\mu\nu}
   &=&-\mbox{\boldmath $j$}_{\nu}
    =i\frac{g}{2}\Big[ 
       \mbox{\boldmath ${\phi}$}^{\dagger} \mbox{\boldmath $\tau$}
       (\hat{D}_{\nu}\mbox{\boldmath ${\phi}$}) 
      -(\hat{D}_{\nu} \mbox{\boldmath ${\phi}$})^{\dagger}
       \mbox{\boldmath $\tau$} \mbox{\boldmath ${\phi}$}\Big],
\label{eqm1}
\end{eqnarray}
\begin{eqnarray}
\partial_{\mu} G_{\mu\nu}
   &=&-k_{\nu} 
    =i\frac{g'}{2}\Big[
       \mbox{\boldmath ${\phi}$}^{\dagger} (\hat{D}_{\nu} 
       \mbox{\boldmath ${\phi}$})
     - (\hat{D}_{\nu} \mbox{\boldmath${\phi}$})^{\dagger}
       \mbox{\boldmath ${\phi}$}\Big]. 
\nonumber 
\end{eqnarray}

Now we choose the following static spherically symmetric ansatz
\begin{eqnarray*}
\mbox{\boldmath$\phi$}=\frac{1}{\sqrt{2}}\rho(r)\xi(\theta,\varphi),
\end{eqnarray*}
\begin{eqnarray*}
\xi=i\left(\begin{array}{cc}
         \sin (\theta/2)\,\, e^{-i\varphi}\\
       - \cos(\theta/2)
      \end{array} \right), 
\hspace{5mm}{\mbox{\boldmath $\hat{\phi}$}} = \xi^{\dagger}
\mbox{\boldmath $\tau$} \xi = - \hat{r},
\end{eqnarray*}
\begin{eqnarray}
\mbox{\boldmath $A$}_{\mu} 
&=& \frac{1}{g} A(r){\mbox{\boldmath $\hat{\phi}$}} \partial_{\mu}t
   +\frac{1}{g}(f(r)-1) {\mbox{\boldmath $\hat{\phi}$}} \times
    \partial_{\mu} {\mbox{\boldmath $\hat{\phi}$}} 
\label{ansatz1},
\end{eqnarray}
\begin{eqnarray}
B_{\mu} &=&-\frac{1}{g'} B(r) \partial_{\mu}t -
            \frac{1}{g'}(1-\cos\theta) \partial_{\mu} \varphi, 
\nonumber
\end{eqnarray}
where $(t,r, \theta, \varphi)$ are the spherical coordinates.
Notice that the apparent string
singularity along the negative z-axis in $\xi$ and $B_{\mu}$ is a pure
gauge artifact which can easily be removed with a hypercharge $U(1)$
gauge transformation. Indeed 
one can easily exociate the string by making the hypercharge $U(1)$ 
bundle non-trivial \cite{Wu}. So {\it the above ansatz describes a most
general spherically symmetric ansatz of a $SU(2) \times U(1)$ dyon}.
Here we emphasize the importance of the non-trivial $U(1)$ 
degrees of freedom to
make the ansatz spherically symmetric. Without the extra $U(1)$ the
Higgs doublet does not allow a spherically symmetric ansatz.
This is because the spherical symmetry for 
the gauge field involves the embedding
of the radial isotropy group $SO(2)$ into the gauge group 
that requires the Higgs field to be
invariant under the $U(1)$ subgroup of $SU(2)$. This is possible
with a Higgs triplet, 
but not with a Higgs doublet \cite{Forg}. In fact, in the
absence of the hypercharge $U(1)$ degrees of freedom, the above ansatz
describes the $SU(2)$ sphaleron which is not spherically
symmetric \cite{Dashen}. The situation changes with the 
inclusion of the extra hypercharge 
$U(1)$ in the standard model, which can compensate the action of
the $U(1)$ subgroup of $SU(2)$ on the Higgs field. 

The spherically symmetric ansatz (\ref{ansatz1}) reduces 
the equations of motion to  
\begin{eqnarray}
\ddot{f} -\frac{f^2-1}{r^2}f  
             =\Big(\frac{g^2}{4}\rho^2 - A^2\Big)f, \nonumber
\end{eqnarray}
\begin{eqnarray}
\ddot{\rho} 
 +\frac{2}{r} \dot{\rho} 
 -\frac{f^2}{2r^2}\rho
 =-\frac{1}{4}(B-A)^2\rho 
 +\lambda\Big(\frac{\rho^2}{2} 
 -\frac{\mu^2}{\lambda}\Big)\rho \nonumber, 
\end{eqnarray}
\begin{eqnarray}
\ddot{A}  
 +\frac{2}{r}\dot{A} 
 -\frac{2f^2}{r^2}A  
 =\frac{g^2}{4}\rho^2(A-B), \label{spher}
\end{eqnarray}
\begin{eqnarray}
\ddot{B}  
 +\frac{2}{r} \dot{B}
 =\frac{g'^2}{4} \rho^2 (B-A). \nonumber 
\end{eqnarray}
The smoothness of the solution requires the following boundary conditions
near the origin,
\begin{eqnarray}
\label{origin}
f    &\simeq& 1+ \alpha_1  r^2, \nonumber \\
\rho &\simeq& \beta_1 r^\delta, \nonumber \\
A    &\simeq& a_1 r,\\
B    &\simeq& b_0 + b_1 r , 
          \nonumber 
\end{eqnarray}
where $\delta =(-1+\sqrt{3})/2$. 
On the other hand asymptotically 
the finiteness of energy requires the following
condition, 
\begin{eqnarray}
\label{infty}
f    &\simeq&  f_1 \exp(-\kappa  r),\nonumber\\
\rho &\simeq& \rho_0 +\rho_1\frac{\exp(-\sqrt{2}\mu r)}{r}, \nonumber \\
A    &\simeq& A_0 +\frac{A_1}{r}, \\
B    &\simeq& A +B_1 \frac{\exp(-\nu r)}{r}, \nonumber 
\end{eqnarray}
where $\rho_0=\sqrt{2\mu^2/\lambda}$, 
$\kappa=\sqrt{(g\rho_0)^2/4 -A_0^2}$,
and $\nu=\sqrt{(g^2 +g'^2)}\rho_0/2$.
Notice  that asymptotically $B(r)$ must approaches to $A(r)$ with an 
exponential damping.

To determine the electric and magnetic charge of the dyon we now
perform the following gauge transformation on (\ref{ansatz1})
\begin{eqnarray}
\xi \longrightarrow \xi'=U \xi = \left(\begin{array}{cc}
0 \\ 1 
\end{array} \right) ,
\label{gauge}
\end{eqnarray}
\begin{eqnarray}
 U=i\left( \begin{array}{cc}
        \cos (\theta/2)& \sin(\theta/2)e^{-i\varphi} \\
        -\sin(\theta/2) e^{i\varphi} & \cos(\theta/2)
\end{array}
\right),\nonumber
\end{eqnarray}
and find that in this unitary  gauge
\begin{eqnarray}
\mbox{ \boldmath $A$}_\mu \longrightarrow
\mbox{ \boldmath $A$}_\mu'=\frac{1}{g} 
\left(
\begin{array}{c}
-f(r)(\sin\varphi\partial_\mu\theta 
    +\sin\theta\cos\varphi \partial_\mu\varphi) \\
f(r)(\cos\varphi\partial_\mu \theta
    -\sin\theta\sin\varphi\partial_\mu\varphi) \\
-A(r)\partial_\mu t  -(1-\cos\theta)\partial_\mu\varphi
\end{array}
\right).\label{unitary}
\end{eqnarray}
So expressing the electromagnetic potential ${\cal A}_\mu$ and the 
neutral potential $Z_\mu$ with the Weinberg angle $\theta_{\rm w}$
\begin{eqnarray}
\left( \begin{array}{cc}
{\cal A}_{\mu} \\  Z_{\mu} 
\end{array} \right)
&=& \left(\begin{array}{cc}
\cos\theta_{\rm w} & \sin\theta_{\rm w}\\
-\sin\theta_{\rm w} & \cos\theta_{\rm w}
\end{array} \right)
\left( \begin{array}{cc}
B_{\mu} \\ A^3_{\mu}
\end{array} \right) \nonumber\\
&=& \frac{1}{\sqrt{g^2 + g'^2}}
\left(\begin{array}{cc}
g & g' \\ -g' & g
\end{array} \right)
\left( \begin{array}{cc}
B_{\mu} \\ A^3_{\mu}
\end{array} \right) , \label{wein}
\end{eqnarray}
we have 
\begin{eqnarray}
\label{eq:Em}
{\cal A}_{\mu} &=& - e \Big( \frac{1}{g^2}A +
    \frac{1}{g'^2}B \Big) \partial_{\mu}t -
  \frac{1}{e}(1-\cos\theta) \partial_{\mu}
 \varphi,  \nonumber \\
Z_{\mu} &=& \frac{e}{gg'}(B-A) \partial_{\mu}t , 
\label{az}
\end{eqnarray}
where $e$ is the electric charge 
\begin{eqnarray*}
e=\frac{gg'}{\sqrt{g^2+g'^2}}=g\sin\theta_{\rm w}.
\end{eqnarray*}
{}From this one has the following electromagnetic charges of the dyon
\begin{eqnarray}
\label{eq:Charge}
q_e &=& {4\pi e}\bigg[
      r^2 \Big(\frac{1}{g^2}\dot{A}+\frac{1}{g'^2}\dot{B} 
      \Big)\bigg] \bigg|_{r=\infty}
     =\frac{4\pi}{e} A_1 \nonumber \\
    &=&\frac{8\pi}{e}\sin^2\theta_{\rm w}\int\limits^\infty_0 f^2 A dr , \\
q_m &=& \frac{4\pi}{e}. \nonumber  
\end{eqnarray}
Also, from the asymptotic condition (\ref{infty})
we conclude that our dyon does not carry any neutral charge,
\begin{eqnarray}
\label{neutral}
Z_e &=&-\frac{4\pi e}{gg'}\Big[ r^2 (\dot{B}-\dot{A})\Big]
        \Big|_{r=\infty}
=0,\nonumber \\ 
Z_m &=& 0,
\end{eqnarray}
which is what one should have expected.

\begin{figure}
\epsfysize=7cm
\centerline{\epsffile{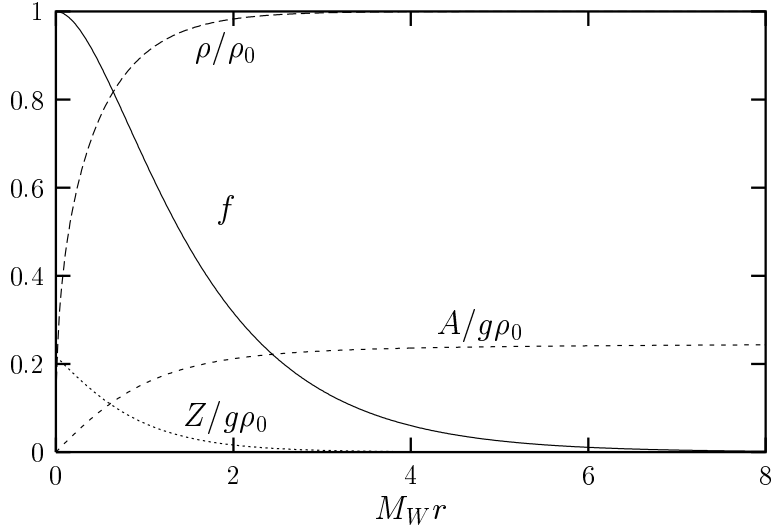}}
\caption{The Cho-Maison dyon solution. Here $Z(r)=B(r)-A(r)$ 
and we have chosen 
$\sin^2\theta_{\rm w}=0.2325$, $\lambda/g^2=M_H^2/2M_W^2=0.5$,
and $A_0=M_W/2$.}
\label{fig1}
\end{figure}

With the boundary conditions one can integrate
(\ref{spher}) and find 
the Cho-Maison dyon solution shown in Fig.\ref{fig1}~\cite{Cho0}.
The regular part of the solution
looks very much like the well-known
Prasad-Sommerfield solution of the Julia-Zee dyon \cite{Julia}.
But there is a crucial difference. 
The Cho-Maison dyon now has a non-trivial
$B-A$, which represents the non-vanishing neutral $Z$ boson content 
of the dyon as shown by (\ref{az}).
To understand the behavior of the solutions, remember
that the mass of the $W$ and $Z$ bosons are given by 
$M_W=g\rho_0/2$ and $M_Z=\sqrt{g^2+g'^2}\rho_0/2$, and the 
mass of Higgs boson is given by $M_H=\sqrt{2}\mu$.
This confirms that $\sqrt{(M_W)^2-(A_0)^2}$ 
and $M_H$ determines the exponential
damping of $f$ and $\rho$, and $M_Z$ determines the 
exponential damping of $B-A$, to their vacuum expectation values
asymptotically.
These are exactly what one would have expected. 

With the ansatz (\ref{ansatz1})
the energy of the dyon is given by
\begin{eqnarray}
E=E_0 +E_1, 
\label{eng1}
\end{eqnarray}
\begin{eqnarray}
E_0=\frac{2\pi}{g^2}\int\limits_0^\infty \frac{dr}{r^2} 
\bigg\{\frac{g^2}{g'^2}
+ (1-f^2)^2\bigg\}, 
\nonumber 
\end{eqnarray}
\begin{eqnarray}
E_1&=&\frac{4\pi}{g^2} \int\limits_0^\infty dr \bigg\{ 
\frac{g^2}{2}(r\dot\rho)^2
+\frac{g^2}{4} f^2\rho^2 +\frac{g^2r^2}{8} (B-A)^2 \rho^2 
+\frac{\lambda g^2r^2}{2}\Big(\frac{\rho^2}{2}-\frac{\mu^2}{\lambda}\Big)^2
\nonumber \\
&& +(\dot f)^2 
+\frac{1}{2}(r\dot A)^2 
+\frac{g^2}{2g'^2}(r\dot B)^2 
+  f^2 A^2
\bigg\}.\nonumber 
\end{eqnarray}
Now with the boundary conditions (\ref{origin}) and (\ref{infty}) one could 
easily find that $E_1$ is finite.
As for $E_0$ we can minimize it with the boundary condition $f(0)=1$, but 
even with this $E_0$ becomes infinite. 
Of course the origin of this 
infinite energy is obvious, 
which is precisely due to the magnetic singularity at
the origin.
This means that one can not predict the mass of dyon. 
It remains arbitrary at the classical level.

\setcounter{section}{3}
\setcounter{equation}{0}
\section*{III. Analytic Solutions}
\indent

At this stage one may ask whether there is any way to make 
the energy of the Cho-Maison solutions finite.
A simple way to make the energy finite is to introduce 
the gravitational interaction~\cite{Bais}.
But the gravitational interaction is not likely remove the singularity
at the origin, and one may still wonder if there is any way to 
regularize the Cho-Maison solutions.
In this section we will show that this is indeed possible,
and discuss how one can construct  
the monopole and dyon solutions explicitly which have not only a finite energy
but also analytic everywhere.

To do this we first notice that a non-Abelian gauge theory in general
is nothing but a special type of an Abelian gauge theory 
which has a well-defined set of charged vector fields as its source.
This must be obvious, but this trivial observation reminds us the fact
that {\em the finite energy non-Abelian monopoles are really nothing but the
Abelian monopoles whose singularity is regularized 
by the charged vector fields.}
{}From this perspective one can try to 
make the energy of the above solutions
finite by introducing additional interactions and/or charged vector fields.
In the followings we will present two ways which 
allow us to achieve this goal along this line,
and construct analytic electroweak monopole and dyon solutions
with finite energy.
\vskip 1em 
\noindent{\large\bf A. Electromagnetic Regularization}
\vskip 1em 

Remember that the origin of the infinite energy of
the Cho-Maison solutions is the magnetic singularity 
of $U(1)_{\rm em}$ at the origin. 
We could try to regularize
this singularity with a judicious choice of an extra 
electromagnetic interaction of the charged vector field
with the Abelian monopole. 
This regularization would provide 
a most economic way to make the energy of the Cho-Maison solution finite, 
because here we could use the already existing $W$ boson without
introducing a new source.

To show that this is indeed possible we first notice 
that in the unitary gauge the Lagrangian (\ref{lag1})
can be written as 
\begin{eqnarray}
{\cal L}
&=&-\frac{1}{4}F_{\mu\nu}^2 -\frac{1}{4} G_{\mu\nu}^2
-\frac{1}{2}|D_\mu W_\nu -D_\nu W_\mu |^2 \nonumber \\
&&+ig F_{\mu\nu} W_\mu^*W_\nu 
+\frac{1}{4}g^2(W_\mu^* W_\nu -W_\nu^* W_\mu)^2 
\nonumber \\
&& -\frac{1}{2}(\partial_\mu \rho)^2
-\frac{1}{4}\rho^2\Big(g^2W_\mu^* W_\mu
 +\frac{1}{2}(g'B_\mu-gA_\mu)^2\Big)
-\frac{\lambda}{2}\Big(\frac{\rho^2}{2}-\frac{\mu^2}{\lambda}\Big)^2,
\label{lag2}
\end{eqnarray}
where
\begin{eqnarray}
W_\mu&=&\frac{1}{\sqrt2}(A^1_\mu+iA^2_\mu),\nonumber  \\
A_\mu&=&A_\mu^3,\nonumber \\
D_\mu W_\nu &=&(\partial_\mu +ig A_\mu)W_\nu .\nonumber 
\end{eqnarray}
This Lagrangian describes the dynamics of two $U(1)$ gauge fields $A_\mu$ and 
$B_\mu$ interacting with a charged vector field $W_\mu$ 
and a real scalar field $\rho$. 
Notice that in the unitary gauge the spherically symmetric ansatz
(\ref{ansatz1}) is written as  
\begin{eqnarray}
\rho&=&\rho(r) \nonumber \\
W_\mu&=& \frac{i}{g}\frac{f(r)}{\sqrt2}e^{i\varphi}
      (\partial_\mu \theta +i \sin\theta \partial_\mu \varphi),
\nonumber \\
A_\mu&=&-\frac{1}{g}A(r)\partial_\mu t 
          -\frac{1}{g}(1-\cos\theta)\partial_\mu \varphi,
\label{ansatz2}\\
B_\mu&=& -\frac{1}{g'} B(r)\partial_\mu t  
          -\frac{1}{g'} (1-\cos\theta) \partial_\mu\varphi,
\nonumber 
\end{eqnarray}
which must be clear from (\ref{unitary}).

To regularize the Cho-Maison dyon, 
we now introduce an extra interaction
${\cal L}_1$ to (\ref{lag2}),
\begin{eqnarray}
{\cal L}_1=i\alpha gF_{\mu\nu}W_\mu^* W_\nu 
               +\frac{\beta}{4}g^2(W_\mu^*W_\nu-W_\nu^*W_\mu)^2,
\label{int1}
\end{eqnarray}
where $\alpha$ and $\beta$ are arbitrary constants.
With this additional interaction the energy of system is given by 
\begin{eqnarray}
E=E_0' +E_1 , 
\label{eng2}
\end{eqnarray}
where now $E_0'$ is given by
\begin{eqnarray}
E_0'=
\frac{2\pi}{g^2}\int\limits_0^\infty
\frac{dr}{r^2}\bigg\{
\frac{g^2}{g'^2}+1-2(1+\alpha)f^2+(1+\beta)f^4
\bigg\}.
\end{eqnarray}
Notice that with $\alpha=\beta=0$, $E_0'$ reduces to $E_0$ and becomes
infinite.
Clearly, for the energy (\ref{eng2}) to be finite, 
$E_0'$ must be free not only from the $O(1/r^2)$ 
singularity  but also $O(1/r)$ singularity at the  origin. 
This requires us to have 
\begin{eqnarray}
\label{cond1}
1+\frac{g^2}{g'^2}-2(1+\alpha) f^2(0)+(1+\beta) f^4(0)=0,  \nonumber 
\end{eqnarray}
\begin{eqnarray}
(1+\alpha)f(0)-(1+\beta) f^3(0) =0 .
\end{eqnarray}
Thus we arrive at  the following condition
in order to have a finite energy 
\begin{eqnarray}
\label{cond3}
\frac{(1+\alpha)^2}{1+\beta}
=1+\frac{g^2}{g'^2}=\frac{1}{\sin^2\theta_{\rm w}} ,
\end{eqnarray}
{}from which we have 
\begin{eqnarray}
f(0)=\frac{1}{\sqrt{(1+\alpha)\sin^2\theta_{\rm w}} }.
\end{eqnarray}
In general $f(0)$ can assume an arbitrary value depending on the choice of 
$\alpha$.
But notice that, except for $f(0)=1$,
the $SU(2)$ gauge field is not 
well-defined at the origin. This means that only when $f(0)=1$,
or equivalently only when $\alpha=\beta$, 
the solution becomes analytic everywhere
including the origin. 
 
\begin{figure}
\epsfysize=7cm
\centerline{\epsffile{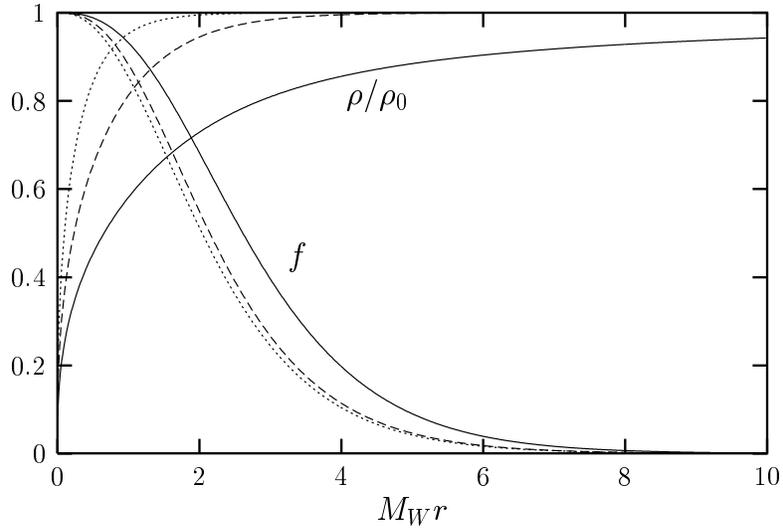}}
\caption{The finite energy electroweak monopole solution
obtained with different values of $\lambda/g^2=0$ (solid line),
$0.5$(dashed line), and $4.5$(dotted line).}
\label{fig2}
\end{figure}

\begin{figure}
\epsfysize=7cm
\centerline{\epsffile{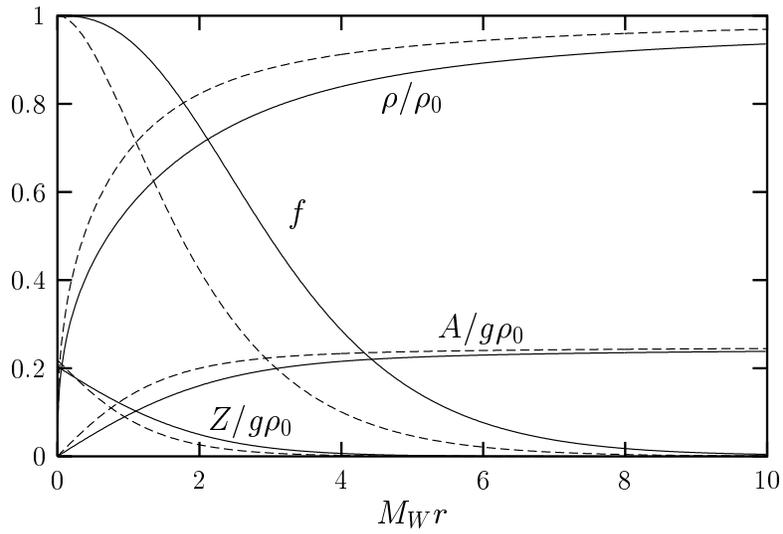}}
\caption{The electroweak dyon solution. 
The solid line represents the finite energy dyon 
and dotted line represents the Cho-Maison dyon, where we have chosen 
$\lambda/g^2=0$ and $A_0=M_W/2$.}
\label{fig3}
\end{figure}

Now, the equations of motion that extremise the energy functional are
given by
\begin{eqnarray}
\ddot f - \frac{(1+\beta)f^2 -(1+\alpha)}{r^2} f 
=\Big(\frac{g^2}{4}\rho^2-A^2 \Big) f ,\nonumber 
\end{eqnarray}
\begin{eqnarray}
\ddot \rho+\frac{2}{r}\dot\rho 
-\frac{f^2}{2r^2}\rho
=-\frac{1}{4}(B-A)^2 \rho 
+\lambda \Big(\frac{\rho^2}{2}
-\frac{\mu^2}{\lambda}\Big)\rho ,\nonumber 
\end{eqnarray}
\begin{eqnarray}
\ddot A +\frac{2}{r} \dot A 
- \frac{2f^2}{r^2} A 
=\frac{g^2}{4}(A-B)\rho^2 ,
\label{eqm3}
\end{eqnarray}
\begin{eqnarray}
\ddot B+\frac{2}{r}\dot B 
=\frac{g'^2}{4}(B-A) \rho^2 .\nonumber 
\end{eqnarray} 
One could integrate this with the boundary conditions 
\begin{eqnarray}
f(0)=1/\sqrt{(1+\alpha)\sin^2\theta_{\rm w}},
~~A(0)=0,~~B(0)=b_0,~~\rho(0)=0,
\nonumber
\end{eqnarray}
\begin{eqnarray}
f(\infty)=0,~~A(\infty)=B(\infty)=A_0,
~~\rho(\infty)=\rho_0.
\end{eqnarray}
Notice that
since $E_1$  contains the term $r^2(B-A)^2\rho^2$,
one must have $A(\infty)=B(\infty)$ to make the energy finite.
Moreover notice that (\ref{eqm3}) is invariant under 
$(A,B)\rightarrow (-A,-B)$. From this symmetry and 
the last two equations of (\ref{eqm3}) one can show that
$B(r)\ge A(r)\ge 0$ for all range~\cite{Yang}.
This tells us that $b_0$ can not be negative.
The results of the numerical integration for the monopole and dyon solution 
are shown in Fig.\ref{fig2} and Fig.\ref{fig3}.
Here we have chosen the experimental value
of $\sin^2\theta_{\rm w}(=0.2325)$, and assumed $f(0)=1$
to guarantee the analyticity of the solution.
{\em It is really remarkable that the finite energy solutions look
almost identical to the Cho-Maison solutions,
even though they no longer have the singularity at the origin
and analytic everywhere.}

Clearly the energy of the above solutions must be of the order of $M_W$. 
Indeed for the monopole the energy can be expressed as 
\begin{eqnarray}
E=\frac{4\pi }{e^2} C(\alpha, \sin^2\theta_{\rm w},\lambda/g^2)M_W
\end{eqnarray}
where $C$ the dimensionless function of 
$\alpha$, $ \sin^2\theta_{\rm w}$, and $\lambda/g^2$. 
With $f(0)=1$ and experimental value $\sin^2\theta_{\rm w}$, $C$
becomes slowly varying function of $\lambda/g^2$ with $C=0.540$ for 
$\lambda/g^2=0$.
This demonstrates that the finite energy solutions can indeed be 
interpreted as the electroweak monopole and dyon, 
and are really nothing but the regularized
Cho-Maison solutions which have a mass of the electroweak scale.

It is interesting to notice that for the monopole solution
we can find the Bogomol'nyi-type energy bound 
if we add an extra term ${\cal L}_2$ in the Lagrangian (\ref{lag2}) 
and (\ref{int1})
\begin{eqnarray}
\label{mass}
{\cal L}_2=-\Big((1+\alpha)^2\sin^2\theta_{\rm w}-\frac{1}{4}\Big)
                   g^2\rho^2W_\mu^* W_\mu .
\label{int2}
\end{eqnarray}
Notice that this amounts to changing the mass of the $W$ boson
from $g\rho_0/2$ to $(1+\alpha)e\rho_0$.
In this case with (\ref{cond3}) 
the energy functional in the Prasad-Sommerfield limit $\lambda=0$
becomes 
\begin{eqnarray}
E&=&\int d^3x \bigg\{
  \frac{1}{2}|D_i W_j -D_j W_i|^2 
 +\frac{1}{2}\Big[ i(1+\alpha)e \epsilon_{ijk}W_j^*W_k 
                  -\frac{g}{2e}\epsilon_{ijk}F_{jk}\Big]^2 
  \nonumber \\
&&\hspace{10mm}
 +\frac{1}{2}(\partial_i \rho)^2 
 +(1+\alpha)^2e^2\rho^2|W_i|^2\bigg\}
 \nonumber \\
&=&\int d^3x \bigg\{ 
 \Big|\epsilon_{ijk} D_j W_k 
      \pm i(1+\alpha)e \rho W_i \Big|^2
\nonumber \\
&&\hspace{10mm}
 +\frac{1}{2}\Big(\partial_i\rho \mp  
                  \epsilon_{ijk}\Big[
                   i(1+\alpha)e W_j^* W_k
                  -\frac{g}{2e}F_{jk}\Big]\Big)^2 \bigg\}
\nonumber \\
&&\mp \frac{1}{e}\int d^3x  
   \partial_i\Big[ \epsilon_{ijk}
    \Big(\frac{g}{2}F_{jk}-i(1+\alpha)e^2 W_j^*W_k\Big)\rho
   \Big] \nonumber \\
&& \pm \frac{1}{e}\int d^3x 
         \frac{g}{2}\epsilon_{ijk}(\partial_i  F_{jk})\rho, 
\end{eqnarray}
where we have used of the fact that $g^2(F_{ij})^2=g'^2(G_{ij})^2$
which follows from  the ansatz (\ref{ansatz2}). 
The last integral gives a delta-function at the origin where
$\rho=0$ so that integral does not contribute. 
And the second integral can be  converted to a surface integral
at the spatial infinity  where the second part 
$ig^2\epsilon_{ijk}\rho W_j^*W_k $ goes faster 
than $O(1/r^2)$, so that
only the first part contributes.
Thus the energy of the monopole is obviously
bounded from below by
\begin{eqnarray}
E \ge \bigg| \frac{1}{e}\int d^3x \partial_i 
           \Big(\frac{g}{2}\epsilon_{ijk} F_{jk}\rho\Big) \bigg|.
\end{eqnarray} 
{}Furthermore  this bound is saturated by the following
equation,
\begin{eqnarray}
 \epsilon_{ijk} D_j W_k
      \pm i (1+\alpha)e\rho W_i=0,\nonumber 
\end{eqnarray}
\begin{eqnarray}
\partial_i\rho \mp
                  \epsilon_{ijk}\Big[
                   i(1+\alpha)e W_j^* W_k
                  -\frac{g}{2e}F_{jk}\Big]=0 ,
\label{self1}
\end{eqnarray}
which is very similar to the well-known 
Bogomol'nyi-Prasad-Sommerfield monopole equation of  the 
Georgi-Glashow model.
The first order differential equation is much more tractable than the 
second order field equation, and also solves it automatically.

Inserting the ansatz (\ref{ansatz2}) into 
(\ref{self1}) we obtain the following Bogomol'nyi-type equation
\begin{eqnarray}
\dot{f}\pm e(1+\alpha)\rho f=0,
\nonumber 
\end{eqnarray}
\begin{eqnarray}
\dot{\rho}\mp
\frac{1}{er^2}\Big(1-(1+\alpha)\sin^2\theta_{\rm w}f^2\Big) =0.
\label{self2}
\end{eqnarray}
Let us consider the upper sign in more detail.
Near the origin, we have 
\begin{eqnarray}
f(r)&\simeq&\frac{1}{\sqrt{(1+\alpha)\sin^2\theta_{\rm w}}}- a r^{l+1}, 
\nonumber \\
\rho(r)&\simeq&\frac{2\sqrt{(1+\alpha)\sin^2\theta_{\rm w}}}{el} a r^l, 
\end{eqnarray}
where $l=(-1+\sqrt{9+8\alpha})/2$. 
On the other hand for large $r$, $f(r)$ approaches to zero exponentially.
Thus the cloud of charged vector fields exists only in the core of monopole.
Also the Higgs field has a exponentially 
decaying piece, with a long-range $1/r$ tail. 
So this solution is again very much like the 
Bogomol'nyi-Prasad-Sommerfield solution of the
Georgi-Glashow model.
\begin{figure}
\epsfysize=7cm
\centerline{\epsffile{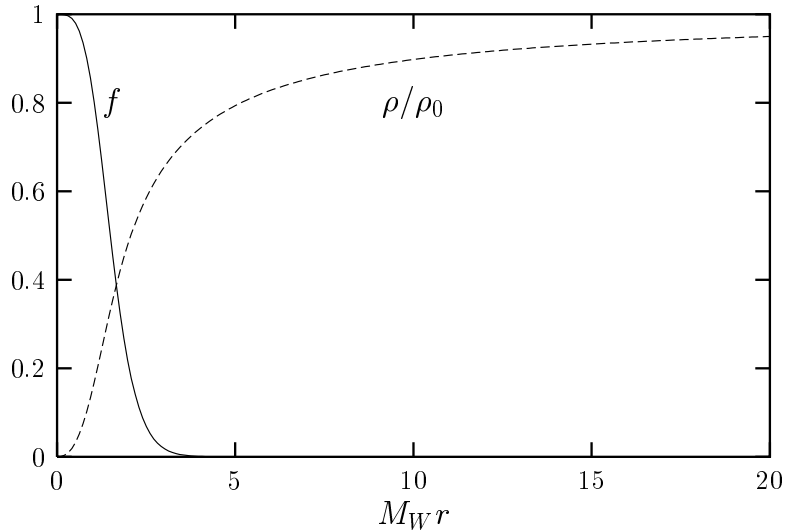}}
\caption{The analytic monopole solution in the Bogomol'nyi-Prasad-Sommerfield
 limit.} 
\label{fig4}
\end{figure}
In fact we can  compare this  with the following 
Bogomol'nyi-Prasad-Sommerfield equation of the Georgi-Glashow
model 
\begin{eqnarray}
\dot{f} \pm e \rho f=0, \nonumber 
\end{eqnarray}
\begin{eqnarray}
\dot{\rho}\mp \frac{1}{er^2}(1-f^2)=0, 
\end{eqnarray}
which was obtained 
with the spherically symmetric ansatz 
\begin{eqnarray}
\mbox{\boldmath$A$}_\mu&=&
       {}\frac{1}{e}(f(r)-1)\hat{r}\times \partial_\mu \hat{r}, 
\nonumber \\
{\bf\Phi}&=&\rho(r)\hat r,
\end{eqnarray}
where $\bf \Phi$ is the Higgs triplet of the Georgi-Glashow model.
This equation has the exact solution 
\begin{eqnarray}
f&=& \frac{e\rho_0 r}{\sinh(e\rho_0r)}\nonumber  ,\\
\rho&=& \rho_0\coth(e\rho_0r)-\frac{1}{er}, 
\end{eqnarray}
but in our case it is not possible to express the solution in terms of 
the elementary functions.
In Fig.\ref{fig4} we have plotted the 
Bogomol'nyi solution of (\ref{self2})
with the experimental value of $\sin^2\theta_{\rm w}$
and $f(0)=1$.

Notice that  the energy of our solution has exactly the same form as the 
Bogomol'nyi-Prasad-Sommerfield monopole,
and is given by
\begin{eqnarray}
E=\frac{4\pi}{e}\rho(\infty) 
=\frac{4\pi}{e^2}\sin^2\theta_{\rm w}M_{W}.
\end{eqnarray}
Obviously the solution is stable energetically
since it is the lowest energy configuration.

\vskip 1em
\noindent{\large\bf B. Embedding $SU(2)\times U(1)$ to $SU(2)\times SU(2)$}
\vskip 1em
As we have noticed the origin of the infinite energy of the Cho-Maison
solutions was the magnetic singularity of $U(1)_{\rm em}$. On the other hand 
the ansatz (\ref{ansatz1}) also suggests that this singularity 
really originates from the magnetic part of the hypercharge
$U(1)$ field $B_\mu$.
So one could try to  to obtain a finite energy monopole solution  
by regularizing this hypercharge $U(1)$ singularity.
This could be done by introducing a hypercharged vector field
to the theory.
A simplest way to do this is, of course, to enlarge the hypercharge $U(1)$
and embed it to another $SU(2)$.

To construct the desired solutions we generalize the 
Lagrangian (\ref{lag2}) by adding the following Lagrangian
\begin{eqnarray}
{\cal L}'&=&-\frac{1}{2}|\tilde D_\mu X_\nu-\tilde D_\nu X_\mu|^2 
+ig' G_{\mu\nu}X_\mu^* X_\nu
+\frac{1}{4}g'^2(X_\mu^* X_\nu -X_\nu^* X_\mu)^2 
\nonumber \\
&&-\frac{1}{2}(\partial_\mu\sigma)^2 
-g'^2\sigma^2 X_\mu^* X_\mu 
-\frac{\kappa}{4}\Big( \sigma^2-\frac{m^2}{\kappa}\Big)^2,
\label{lag4}
\end{eqnarray}
where
$X_\mu$ is a hypercharged vector field, $\sigma$ is a Higgs field, and 
$\tilde D_\mu X_\nu=(\partial_\mu +ig' B_\mu) X_\nu$. 
Notice that, if we introduce a hypercharge $SU(2)$ gauge field 
$\mbox{\boldmath$B$}_\mu$ and a scalar triplet ${\bf\Phi}$ and 
identify
\begin{eqnarray}
X_\mu &=&\frac{1}{\sqrt{2}}(B_\mu^1+i B_\mu^2)\nonumber ,\\  
B_\mu &=& B_\mu^3 \nonumber ,\\
{\bf\Phi}&=& (0,0,\sigma),
\end{eqnarray}
the above Lagrangian becomes identical to 
\begin{eqnarray}
{\cal L}'= -\frac{1}{2}(\tilde D_\mu {\bf\Phi})^2
-\frac{\kappa}{4}\Big({\bf\Phi}^2-\frac{m^2}{\kappa}\Big)^2
-\frac{1}{4}\mbox{\boldmath$G$}_{\mu\nu}^2, 
\end{eqnarray}
in the unitary gauge.
This clearly  shows that 
Lagrangian (\ref{lag4}) is nothing but the embedding of the 
hypercharge $U(1)$ to an $SU(2)$ Georgi-Glashow model.

{}From (\ref{lag2}) and (\ref{lag4}) 
one has the following equations of motion
\begin{eqnarray}
\partial_\mu(\partial_\mu \rho)=\frac{1}{2} g^2 W_\mu^* W_\mu \rho
+\frac{1}{4}(g'B_\mu -g A_\mu)^2 \rho
+\lambda\Big(\frac{\rho^2}{2}+\frac{\mu^2}{\lambda}\Big)\rho
,\nonumber
\end{eqnarray}
\begin{eqnarray}
D_\mu (D_\mu W_\nu -D_\nu W_\mu)=ig F_{\mu\nu}W_\mu
-g^2 W_\mu (W_\nu W_\mu^* -W_\nu^* W_\mu)
+\frac{1}{4}g^2\rho^2 W_\nu
,\nonumber 
\end{eqnarray}
\begin{eqnarray}
\partial_\mu F_{\mu\nu}&=& 
\frac{1}{4}g\rho^2(gA_\nu-g'B_\nu)
+ig\Big( W_\mu^*(D_\mu W_\nu -D_\nu W_\mu)
        -(D_\mu W_\nu^* -D_\nu W_\mu^*)W_\mu\Big)
\nonumber \\
&&+ig \partial_\mu (W_\mu^* W_\nu -W_\nu^* W_\mu)
,\nonumber 
\end{eqnarray}
\begin{eqnarray}
\partial_{\mu} G_{\mu\nu}&=& 
\frac{1}{4}g'\rho^2 (g'B_\nu-gA_\nu) 
  +ig'\Big( X_\mu^*(\tilde D_\mu X_\nu -\tilde D_\nu X_\mu)
             -(\tilde D_\mu X_\nu^* -\tilde D_\nu X_\mu^*)X_\mu\Big) 
\nonumber \\
&&    + ig'\partial_\mu (X_\mu^* X_\nu-X_\nu^* X_\mu),
\nonumber 
\end{eqnarray}
\begin{eqnarray}
\partial_\mu(\partial_\mu \sigma)=2g'^2 X_\mu^* X_\mu \sigma 
+\kappa\Big( \sigma^2-\frac{m^2}{\kappa}\Big)\sigma ,\nonumber 
\end{eqnarray}
\begin{eqnarray}
\tilde D_\mu (\tilde D_\mu X_\nu -\tilde D_\nu X_\mu)
  =ig'G_{\mu\nu} X_\mu  
  -g'^2X_\mu (X_\mu^* X_\nu-X_\nu^* X_\mu)
  +(g')^2\sigma^2 X_\nu 
\label{eom5}
\end{eqnarray}
Now for a static spherically symmetric ansatz
we choose (\ref{ansatz2}) and assume
\begin{eqnarray}
\sigma &=&\sigma(r), \nonumber \\
X_\mu &=&\frac{i}{g'}\frac{h(r)}{\sqrt{2}}e^{i\varphi} (\partial_\mu \theta
+i\sin\theta\partial_\mu \varphi).
\label{ansatz3}
\end{eqnarray}
With the spherically symmetric  ansatz 
(\ref{eom5}) is reduced to  
\begin{eqnarray}
\ddot{f} - \frac{f^2-1}{r^2}f = 
              \Big(\frac{g^2}{4}\rho^2 - A^2\Big)f, \nonumber
\end{eqnarray}
\begin{eqnarray}
\ddot{\rho} + \frac{2}{r} \dot{\rho} - \frac{f^2}{2r^2}\rho
  =- \frac{1}{4}(B-A)^2\rho + \lambda\Big(\frac{\rho^2}{2} -
   \frac{\mu^2}{\lambda}\Big)\rho \nonumber, 
\end{eqnarray}
\begin{eqnarray}
\ddot{A} + \frac{2}{r}\dot{A} -\frac{2f^2}{r^2}A = \frac{g^2}{4}
   \rho^2(A-B), \label{eq:Spher}
\end{eqnarray}
\begin{eqnarray}
\ddot h -\frac{h^2-1}{r^2} h =(g'^2\sigma^2-B^2) h ,
\end{eqnarray}
\begin{eqnarray}
\ddot\sigma +\frac{2}{r}\dot\sigma -\frac{2h^2}{r^2} \sigma
= \kappa\Big(\sigma^2-\frac{m^2}{\kappa}\Big)\sigma , \nonumber 
\end{eqnarray}
\begin{eqnarray}
\ddot{B} + \frac{2}{r} \dot{B}- \frac{2h^2}{r^2} B
 =  \frac{g'^2}{4} \rho^2 (B-A). \nonumber  
\label{eom4}
\end{eqnarray}

\begin{figure}
\epsfysize=7cm
\centerline{\epsffile{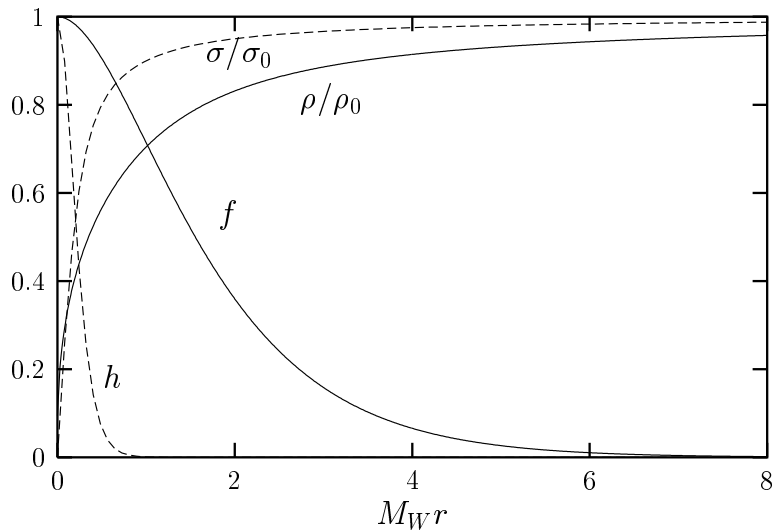}}
\caption{The $SU(2)\times SU(2)$ monopole solution, where 
the dashed line represents hypercharge part which 
describes Bogomol'nyi-Prasad-Sommerfield solution.}
\label{fig5}
\end{figure}

Notice that the energy of the above 
configuration is given by
\begin{eqnarray}
E=E_W +E_X,
\end{eqnarray}
\begin{eqnarray}
E_{W}&=&
\frac{4\pi}{g^2}\int\limits_0^\infty dr 
\bigg\{ 
(\dot f)^2 +\frac{(1-f^2)^2}{2r^2} +\frac{1}{2}(r\dot A)^2
+f^2A^2   \nonumber \\
&&
+\frac{g^2}{2}(r\dot\rho)^2
+ \frac{g^2}{4} f^2\rho^2
+\frac{g^2r^2}{8}(B-A)^2\rho^2
+\frac{\lambda g^2r^2}{2}\Big( \frac{\rho^2}{2}-\frac{\mu^2}{\lambda}\Big)^2
\bigg\}\nonumber \\
&=&\frac{4\pi}{g^2}C_1(\lambda/g^2) M_W ,
\nonumber 
\end{eqnarray}
\begin{eqnarray}
E_X&=&\frac{4\pi}{g'^2}\int\limits_0^\infty dr\bigg\{ 
(\dot h)^2 
+\frac{(1-h^2)^2}{2r^2}
+\frac{1}{2}(r\dot B)^2
+h^2B^2 \nonumber \\
&& 
+\frac{g'^2}{2}(r\dot\sigma)^2
+g'^2 h^2\sigma^2
+\frac{\kappa g'^2r^2}{4}(\sigma^2-\sigma_0^2)^2
\bigg\}\nonumber \\
&=&\frac{4\pi }{g'^2} C_2(\kappa/g'^2)M_X ,
\nonumber
\end{eqnarray}
where $M_W=g\rho_0/2$,and  $M_X=g'\sigma_0=g'\sqrt{m^2/\kappa}$.
The boundary conditions for a regular 
field configuration can be chosen as 
\begin{eqnarray}
f(0)=h(0)=1,~~ A(0)=B(0)=\rho(0)=\sigma(0)=0, \nonumber 
\end{eqnarray}
\begin{eqnarray}
f(\infty)=h(\infty)=0,
~~A(\infty)=B(\infty)=A_0,~~\rho(\infty)=\rho_0,~~\sigma(\infty)=\sigma_0.
\label{bound3}
\end{eqnarray}
Notice that the origin of the condition $B(0)=0$ is the same as $A(0)=0$.
With the boundary condition (\ref{bound3}) one may try to find the desired 
solution. From the physical point of view one could assume $M_X \gg M_W$,
where $M_X$ is an intermediate scale which lies somewhere between
the grand unification scale and the electroweak scale. 
Now, let $A=B=0$ for simplicity. Then (\ref{eom4}) decouples to describes
two independent systems so that the monopole solution has two
cores, the one with the size $O(1/M_W)$ and the other with 
the size $O(1/M_X)$. With $M_X=10M_W$  
we obtain the 
solution shown in Fig.\ref{fig5}
in the limit $\lambda=\kappa=0$.   
In this limit we find  
$C_1=1.946$ and $C_2=1$ so that the  energy of the solution  
is given by
\begin{eqnarray}
E=\frac{4\pi}{e^2}\Big( \cos^2\theta_{\rm w}
          +0.195\sin^2\theta_{\rm w}\Big)M_X.
\end{eqnarray}
Clearly the solution describes 
the Cho-Maison monopole whose singularity is regularized by  
a Prasad-Sommerfield monopole of the size $O(1/M_X)$.

It must be emphasized that even if  the energy of the monopole
is fixed by the intermediate scale, the monopole should be interpreted
as an electroweak monopole. To see this notice that the size
of the monopole is fixed by the electroweak scale. Furthermore from the 
outside the monopole looks exactly the same as the Cho-Maison
monopole. Only the inner core is regularized by the hypercharged vector field.
This justifies it as an electroweak monopole.

\setcounter{section}{4}
\section*{IV. Conclusions}
\indent

In this paper we have discussed two ways to regularize the Cho-Maison 
monopole and dyon solutions of the Weinberg-Salam model, 
and explicitly constructed 
genuine finite energy electroweak monopole and dyon solutions
which are analytic everywhere including the origin.
The finite energy solutions are obtained with a simple modification
of the interaction of the $W$ boson or with the embedding of 
the hypercharge $U(1)$ to a compact $SU(2)$.
It has generally been believed that the finite
energy monopole must exist only at the grand unification
scale~\cite{Dokos}. But our result tells that this belief is 
unfounded, and endorses the existence of a totally new class of electroweak
monopole whose mass is much smaller than the monopoles of the 
grand unification.
Obviously the electroweak monopoles are topological solitons which 
must be stable. 

Strictly speaking the finite energy solutions are not the solutions of the
Weinberg-Salam model, because their existence requires a generalization of the 
model. But from the physical point of view there is no doubt that they should
be interpreted as the electroweak monopole and dyon, because 
they are really nothing but the regularized Cho-Maison solutions whose size is 
fixed at the electroweak scale. In spite of the fact that the Cho-Maison
solutions are obviously the solutions of the Weinberg-Salam model
one could try to object them as the electroweak dyons 
under the presumption that the 
Cho-Maison solutions could be regularized only at the grand unification scale.
Our work shows that 
this objection is groundless, 
and assures that 
it is not necessary for us to go to the 
grand unification scale to make the energy of the Cho-Maison solutions finite.
This really reinforces the Cho-Maison dyons as the 
electroweak dyons which must be taken seriously.

Another important aspect of our result is that, unlike the Dirac monopole,
the magnetic charge of the electroweak monopoles must satisfy
the Schwinger quantization condition $q_m=4\pi n/e $.
The electroweak unification simply forbids the electromagnetic monopole
with $q_m=2\pi/e$.
This is because the $U(1)_{\rm em}$ is defined with the $U(1)$ subgroup
of $SU(2)$, which affects the magnetic charge.
So within the framework of the electroweak unification
the unit of the magnetic charge must be $4\pi/e$. 

We close with the following remarks:

\noindent{1)} The electromagnetic regularization of the Dirac monopole 
with the charged vector fields is nothing new.
In fact it was this regularization which made the
energy of the 't Hooft-Polyakov monopole finite.
Furthermore it has been known that
the 't Hooft-Polyakov monopole 
is the only analytic solution (with $\alpha=\beta=0$)
which one could obtain with this technique~\cite{klee}.
What we have shown in this paper is that the same
technique also works to regularize the Cho-Maison solutions, but
with nonvanishing $\alpha$ and $\beta$.

\noindent{2)} The introduction of the additional interactions (\ref{int1})
and (\ref{int2}) to the Lagrangian (\ref{lag1}) could spoil the 
renormalizability of the Weinberg-Salam model (although this issue has to 
be examined in more detail). How serious would this offense, however,
is not clear at this moment. Here we simply notice that
the introduction of a non-renormalizable interaction 
(like a gravitational interaction)  has been an acceptable practice
to study finite energy  classical solutions.

\noindent{3)} The embedding of the electroweak $SU(2)\times U(1)$ to
a larger $SU(2)\times SU(2)$ or 
$SU(2)\times SU(2)\times U(1)$ could 
naturally arise in the left-right symmetric grand unification
models, in particular in the $SO(10)$ grand unification, although
the embedding of the hypercharge $U(1)$ to a compact $SU(2)$
may turn out to be too simple to be realistic.
Independent of the details, however, our discussion suggests  
that the electroweak monopoles at an intermediate scale
$M_X$ could be  possible in a realistic grand unification.

Certainly
the existence of the finite energy electroweak monopoles should have important
physical implications~\cite{Preskill}. 
We will discuss on the physical implications of the electroweak monopoles
separately.
\section*{Acknowledgments}
\indent

It is a pleasure to thank C.K. Lee for discussions.
The work is supported in part by the Ministry of Education and by the Korean
Science and Engineering Foundation.

\end{document}